\documentclass[a4paper]{article}
\usepackage{epsfig,amssymb,amsmath}

\def\beq{\begin{eqnarray}}
\def\eeq{\end{eqnarray}}
\def\bsp{\begin{split}}
\def\esp{\end{split}}
\def\lra{\longrightarrow}
\def\ra{\rightarrow}
\def\inl{\left\langle}
\def\inr{\right\rangle}

\newcommand{\mc}[1]{{\cal #1}}
\newcommand{\mbold}[1]{\mbox{\boldmath{\ensuremath{#1}}}}

\begin{document}

\title{\textbf{The Weyl Curvature Conjecture}}
\author{\textbf{\O yvind Gr\o n}$^{1,2}$\thanks{Oyvind.Gron@iu.hio.no}
~\textbf{and Sigbj\o rn Hervik}$^3$\thanks{S.Hervik@damtp.cam.ac.uk}
\\ \\
$^1$\small Department of Physics, University of Oslo, PO Box 1048 Blindern,\\
\small N-0316 Oslo, Norway \\ \\
$^2$\small Oslo College, Faculty of Engineering, Cort Adelers
gt. 30,\\ \small N-0254
Oslo, Norway
\\ \\
$^3$\small DAMTP,
Centre for Mathematical Sciences,
Cambridge University, \\
\small Wilberforce Rd.,
Cambridge CB3 0WA, UK}
\date{\small \today}
\maketitle

\begin{abstract}
In this paper we review Penrose's Weyl curvature conjecture which
states that the concept of gravitational entropy and the Weyl tensor
is somehow linked, at least in a cosmological setting. We give a description of
 a certain entity constructed from the Weyl tensor, from the very
early history of our universe until the present day. Inflation is an important mechanism in
our early universe for homogenisation and isotropisation, and thus it
must cause large  effects upon the evolution of the gravitational
entropy. Therefore the effects from inflationary fluids and a
cosmological constant are studied in detail.  
\end{abstract}

\section{The arrow of time and gravitational entropy in the context of a cosmology}
     There is a strange omission in the traditional version of the second law of thermodynamics 
(SLT). It does not take gravity into account. However the existence of the arrow of time is 
usually explained with reference to SLT. And most discussions of the origin of the arrow of 
time appeal ultimately to the initial condition and the evolution of the universe. Gravity plays 
an essential role for this evolution.

   As pointed out by P.C.W. Davies \cite{davies74,davies83} there seems to be a paradox that the material contents of 
the universe began in a condition of thermodynamic equilibrium, whereas the universe today 
is far from equilibrium. Hence the thermodynamic entropy has been reduced in conflict with 
SLT which says that the entropy of the universe is an increasing function of cosmic time. This 
is due to the tendency of self-gravitating systems irreversibly to
grow inhomogeneous. 

   In order to include this gravitational effect into a generalized version of SLT, one has to 
define a gravitational entropy. Several tentative definitions 
have been given based on Penrose's Weyl curvature hypothesis \cite{penrose77,penrose,penrose94}. 

   The Weyl curvature tensor vanishes identically in the homogeneous and isotropic 
Friedmann-Robertson-Walker universe models. It was suggested by Penrose \cite{penrose77,penrose,penrose94} to use this 
tensor as a measure of inhomogeneities of the universe models. The Weyl curvature scalar is 
expressed by scalars constructed from the Riemann curvature tensor as
follows
\beq
C^{\alpha\beta\gamma\delta}C_{\alpha\beta\gamma\delta}=R^{\alpha\beta\gamma\delta}R_{\alpha\beta\gamma\delta}-2R^{\alpha\beta}R_{\alpha\beta}+\frac
13 R^2
\eeq
	 
   Penrose has given several formulations of the conjecture. A version that emphasizes the 
entropy aspect was given in ref. \cite{penrose77} p.178.
It seems that in some way the Weyl tensor gives a measure of the entropy in the space-
time geometry. The initial curvature singularity would then be one with large Ricci 
tensor and vanishing Weyl tensor (zero entropy in the geometry); the final curvature 
singularity would have Weyl tensor much larger than Ricci tensor (large entropy in the 
geometry).

In order to quantify Penrose's conjecture Wainwright and collaborators
\cite{wa,gw} have suggested that the quantity
\beq
P^2=\frac{C^{\alpha\beta\gamma\delta}C_{\alpha\beta\gamma\delta}}{R^{\mu\nu}R_{\mu\nu}}
\label{defP}\eeq
may represent a ``gravitational entropy'', at least in a cosmological
context with a non-vanishing Ricci-tensor. 

We could of course ask ourself whether the entropic behavior of the
Weyl curvature invariant (or an invariant composed thereof) results
because the Weyl tensor is \emph{directly related} to the
gravitational entropy, or whether it is only a side-effect
because the universe evolves towards a state of maximal
entropy. Hence, even though the results may
show that the Weyl curvature invariant has an entropic behaviour, it
is by no means a proof that the Weyl tensor should be identified as
the ``gravitational entropy''.

\section{The behaviour of the Weyl tensor}
In an earlier article \cite{wcc1} we discussed the Weyl curvature
tensor in a homogeneous but anisotropic model as well as in an inhomogeneous model. Our
survey will be summarized in this and the next sections, as well as bringing new
arguments into the discussion.

As a basis for our study, we used two different models which induce
Weyl curvature effects in two conceptually different ways. One was the
anisotropic Bianchi type I model, the other was the inhomogeneous
Lema\^itre-Tolman model. The Weyl curvature conjecture has been
investigated in the Szekeres cosmological model, that generalize the
Lema\^itre-Tolman model, by W.B. Bonnor \cite{bonnor86}. Inhomogeneous modes are in general
local modes for general relativity while anisotropic modes are global
modes. The global
topology and geometry of our universe has significant consequences for
the possibilities for anisotropic modes
\cite{Kodama1,BK1,BK2,Kodama2}. The role of inhomogeneous modes can be
considered as more local in origin, but the effect on the global
geometry and topology are not known in detail. Hence, it is 
important to study both an inhomogeneous model as well as a
homogeneous model to get a more complete picture
of the behaviour of the Weyl tensor in generic cosmological models. 

\subsection{The Bianchi type I model}
The Bianchi type I model is the simplest anisotropic
generalisation of the FRW models. It has flat spatial sections, and
hence, can be a good candidate for the universe we live in. The metric
for this model can be written as
\beq
ds^2=-dt^2+e^{2\alpha}\left[e^{2{\beta}}\right]_{ij}dx^idx^j
\eeq
where
${\beta}=\text{diag}(\beta_++\sqrt{3}\beta_-,\beta_+-\sqrt{3}\beta_-,
-2\beta_+)$. In \cite{sigBI} we solved the Einstein field equations
for dust and a cosmological constant $\Lambda$. By introducing the volume element $v=e^{3\alpha}$, the
Einstein field equations gives the following equation for $v$:
\beq
\dot{v}^2=3\Lambda v^2+3Mv+A^2
\eeq
where $M$ is the total mass of the dust, and $A$ is an anisotropy
parameter. The equations for $\beta_{\pm}$ are in terms of the volume element
\beq
\dot{\beta}_{\pm}=\frac{a_{\pm}}{3v}
\eeq 
where the constants $a_{\pm}$ are related to $A$ via
$a_+^2+a_-^2=A^2$. It is useful to define an angular variable $\gamma$ by
$a_+=A\sin(\gamma-\pi/6)$ and $a_-=A\cos(\gamma-\pi/6)$.
An
important special case of the Bianchi type I is the Kasner vacuum
solutions. The Kasner solutions are characterised only by the angular
variable $\gamma$.  These solutions have a Weyl scalar
\beq
\left(C^{\alpha \beta \gamma \delta}C_{\alpha \beta \gamma \delta}\right)_{I}=
\frac{16}{27}\frac{A^4}{v^4}\left(1-2z\cos 3\gamma+z^2\right)
\label{weylBtI}\eeq
where $z=\dot v/A$ and  $z=1$ for the Kasner solutions which have
vanishing Ricci tensor. So in a sense, the Kasner solutions are the
anisotropic counterpart to the inhomogeneous Schwarzschild solution. 

The Weyl tensor will decay as the volume expands, even at late
times. One should expect that this entity would increase
if it represents gravitational entropy. But is decreases monotonically, and hence, it
is doubtful that it is the correct measure in these models. 

 Inserting dust and a
cosmological constant will not have a significant effect on this
decreasing behaviour. If $M$ is the mass of the dust inside the volume
$v$, then eq. (\ref{weylBtI}) still  holds but with
\beq
z=\sqrt{1+\frac{3M v}{A^2}+\frac{3\Lambda
v^2}{A^2}}. 
\eeq
Hence, except for the special case\footnote{The special
case of $\gamma=0$ is somewhat interesting. For $\gamma=0$ it can be
shown that the Kasner solution is just a special part of Minkowski
spacetime, which has no singularities. It seems a bit odd that for
any $\gamma\neq 0$ we will have an initial singularity, while for
$\gamma=0$ we do not have one. However, if we compactify the spatial
sections this oddity disappears. The model has a singularity at $v=0$
even for the case $\gamma=0$!} $\gamma=0$, the Weyl tensor 
diverges as $v^{-4}$ as $v\lra 0$.  

For the Bianchi type I universe models the entity $P^2$
defined in eq. (\ref{defP}), turns out
to be 
\beq
 (P^2)_{I} =\frac{4}{27}\frac{A^4}{v^2}\frac{1+z^2-2z\cos3\gamma}{M^2+2\Lambda Mv
+4\Lambda^2v^2}.
\eeq
Also this measure for the Weyl entropy diverges as $v\lra
0$.  

\subsection{The Lema\^itre-Tolman models}
Let us now consider the inhomogeneous Lema\^itre-Tolman (LT) models. The
line element for the LT models can be written as 
\beq\label{LTmetric}
ds^2=-dt^2+Q^2dr^2+R^2(d\theta^2+\sin^2\theta d\phi^2)
\eeq
where $Q=Q(r,t)$ and $R=R(r,t)$. In \cite{sigLT} we used this
model and investigated the solutions of
 the Einstein field equations where the spacetime contains a cosmological 
constant $\Lambda$ and dust. The
equations then turn into 
\[ R'=FQ\]
where $F=F(r)$ is an arbitrary function, and
\beq\label{meq}
\frac{1}{2}R\dot{R}^2+\frac{1}{2}(1 -F^2)R-\frac{\Lambda}{6}R^3=m.
\label{masseq}\eeq
The function $m=m(r)$ is given by the integral  
\beq
m(r)=\int^r_0 4\pi\rho R^2 R'dr,
\eeq
where prime denotes derivative with respect to $r$ and $\rho$ is the
dust density. Hence, $m(r)$ can be interpreted as the total mass 
of the dust inside the spherical shell of coordinate radius
$r$. Inverting the equation for $m(r)$, the dust density can be
written in terms of $R$ and $m(r)$:
\beq
4\pi\rho=\frac{m'}{R^2R'}
\label{rhoeq}\eeq

It is also useful to define the \emph{mean dust density function}
$\bar{\rho}(r)$ by the relation
\beq
m(r)= \frac{4}{3}\pi\bar{\rho}R^3.
\label{rhobareq}\eeq
Interestingly, the Weyl curvature scalar can now be written quite elegantly as
\beq
\left(C^{\alpha \beta \gamma\delta}C_{\alpha \beta \gamma\delta}\right)_{LT}=
\frac{16^2}{3}\pi^2(\bar{\rho}-\rho)^2.
\label{rhoWeyl}\eeq
This relation provides us with a physical interpretation of the Weyl
scalar in the LT models. It is just the difference
between the mean dust density and the actual dust density. Also, the
Weyl tensor is everywhere zero, if and only if $\bar{\rho}=\rho$. In
this case the LT models turn into the FRW models with homogeneous
spatial sections. The FRW models are conformally flat, hence they have
a zero Weyl tensor. 

Let us return to the equations of motion for the LT model.  The mass
equation (\ref{masseq}) can be written as an ``energy'' equation:
\beq
\frac{1}{2}\dot{R}^2+V(r,R)=E(r)
\label{energyeq}\eeq
where the ``potential'' $V$ and the ``energy'' $E$ are given by
\beq V&= &-\frac{m}{R}-\frac{\Lambda}{6}R^2 \nonumber \\
 E&=&-\frac{1}{2}(1-F^2)
\eeq
\begin{figure} 
\centering
\epsfig{figure=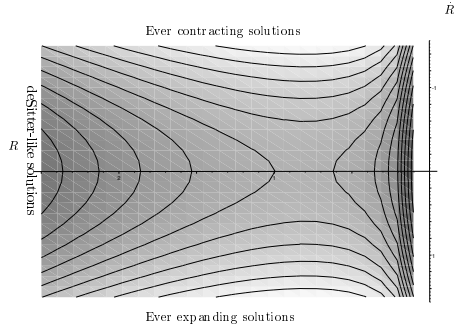, height=5cm}
\caption{Level curves of $E(R, \dot{R})=\frac{1}{2}\dot{R}^2-\frac{1}{5R}-\frac{1}{5}R^2$} \label{levelcurves}
\end{figure}
This energy equation may be integrated and solved exactly. A summary of the results and a qualitative description of the physical meaning of the solutions is given in ~\cite{Kra97}. The solutions may be written in terms of the Weierstrass' elliptic functions ~\cite{Zecca}. The actual expressions are not very informative unless the reader has massive knowledge of these elliptic functions. However a lot of qualitative information can be extracted from simple classical considerations. 

The classical solutions will move on level curves of the energy function {$E(R, \dot{R})=\frac{1}{2}\dot{R}^2-\frac{m}{R}-\frac{\Lambda}{6}R^2$}, since the total energy $E$ is independent of $t$. In figure \ref{levelcurves} the level curves of a typical energy function are drawn. If $\Lambda>0$ there will exist a saddle-point of the energy function. This saddle point will be at $R=\left(\frac{3m}{\Lambda}\right)^{\frac{1}{3}}, \dot{R}=0$ where the energy function will have the value { $ E_s=-\frac{1}{2}(9m^2\Lambda)^{\frac{1}{3}}$}. The saddle point solution is  static, and is as a matter of fact the Einstein static universe\footnote{This is easily seen if we define the new radial variable to be $R$. This can be done since we have to assume that $m'(r)>0$ on physical grounds. In the case $m'(r)=0$, the metric become degenerate.}.
If $E<E_s$, the solutions fall into two distinct classes:
\begin{enumerate}
\item{} Schwarzschild-like solutions: These solutions expands, but they do not possess enough energy to escape the gravitational collapse, so they end as  black holes. If $m$ is constant these solutions are those of a Schwarzschild black hole in Lema\^itre coordinates. 
\item{} de Sitter-like solutions: solutions where the universe evolves
approximately as that of de Sitter solutions with positively curved hypersurfaces.  
\end{enumerate}
If $E>E_s$ the (test) matter has enough energy to escape the
gravitational collapse (expanding solutions) or enough energy to
prevent the gravitational repulsion from the cosmological constant
(contracting solutions).

Near the initial singularity, both $\bar{\rho}$ and $\rho$
diverge. Unless they are identically equal, the Weyl tensor will
diverge near the initial singularity. Explicitly we have near the
initial singularity 
\beq
4\pi(\bar{\rho}-\rho)=\frac{3m}{R^3}\frac{2t'_0}{2t'_0-\frac{m'}{m}(t-t_0)}
\eeq
where we have used that near the initial singularity, we can
approximate the solutions with
\begin{equation} 
R\approx \left(\frac{9}{{2}}m\right)^{\frac{1}{3}}(t-t_0(r))^{\frac{2}{3}}.
 \label{smallimit}\end{equation}
The free function $t_0(r)$ is the big bang time. To avoid intersecting world-lines we have to assume
$t'_0<0$\cite{miller}, thus $(\bar{\rho}-\rho)>0$. The Weyl scalar will
diverge as $R\lra 0$ unless $t'_0=0$. Also the entity $P$ will diverge
in general near the initial singularity. $P$ is given
by
\begin{equation}
\left(P^2\right)_{LT} =\frac{4}{3}\frac{\left(\frac{\bar{\rho}}{\rho}-1\right)^2}{\left(
\frac{\Lambda}{4\pi\rho}\right)^2+\left(\frac{\Lambda}{4\pi\rho}\right)+1}
\end{equation}
which near the initial singularity can be approximated by
\beq
\left(P^2\right)_{LT}\propto \left(\frac{2mt'_0}{m'(t-t_0)}\right)^2.
\eeq
Hence, $P$ diverges as $t\lra t_0$ unless $t'_0=0$, i.e. unless the big
bang is homogeneous.

This does not prove to be a
very promising behaviour for the WCC. The curvature scalars diverge
near the initial singularity, and hence, come in conflict with the
WCC. But let us analyse the situation in the LT a bit more carefully. As
the universe expands, both $\bar{\rho}$ and $\rho$ will decrease. They
both decrease from an infinite value at the initial singularity. As
the universe expands the value of $\bar{\rho}$ decrease as $t^{-2}$
while $\rho$ decreases as $t^{-1}$ close to the initial
singularity. Hence, the value of $\bar{\rho}$ approaches the value of
$\rho$ by a factor or $t^{-1}$, but since they both diverge, the Weyl
tensor and $P$ diverge as well. 

\begin{figure} 
\centering
\epsfig{figure=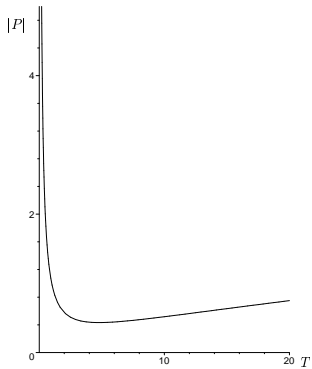, height=5cm}
\caption{The plot of the function $|P|=T^{-1}-\frac 1{10} T^{-\frac
13}+\frac 1{10}T^{\frac 23}$. } \label{Pperturb}
\end{figure}

However, even though R. Penrose and S. Hawking \cite{HP} showed that
according to the general theory of relativity the big bang must have
started in a singularity, it has later been emphasized that the
physical universe must obey not only the general relativistic laws of
nature, but also the quantum mechanical laws. Hence, the initial
singularity is a fiction which not corresponds to physical reality. The
classical laws can be applied only after the Planck time. 
We should therefore study the behaviour of $P^2$ not only in the limit
$t\lra 0$, but rather at a very small cosmic time. Choosing the origin
of cosmic time $T$ at $t=t_0$ we introduce $T=t-t_0(r)$. Hence close
to $T=0$
\beq
|P|=2\frac{m}{m'}\frac{|t_0'|}{T}
\eeq
which implies that $\left[\partial |P|/\partial T\right]_{T\lra
0}<0$. Bonnor \cite{Bon85} considered the model with $1-F^2>0$ and a vanishing
cosmological constant, and found the opposite result. However he
restricted his investigation to models with homogeneous initial
singularity, i.e. $t'_{0}=0$. As we will show this is an exceptional
case. The behaviour of more general models with $t'_{0}\neq 0$ is
different. To see what is actually happening if we push $t'_{0}$
towards zero, we utilize that close to $T=0$ the models with $1-F^2>0$
and $\Lambda=0$ behave according to 
\beq
R^3=\frac 92 mT^2\left(1-\epsilon(r)T^{\frac 23}\right)
\label{smallR3}\eeq
where $\epsilon(r)$ is assumed to be a small function of $r$. 
From eqs. (\ref{rhoeq}), (\ref{rhobareq}) and (\ref{rhoWeyl}) with
$\Lambda=0$ we get
\beq
|P|=\frac{2}{\sqrt{3}}\left(3\frac{mR'}{m'R}-1\right).
\eeq
Using (\ref{smallR3}) we obtain 
\beq
|P|=\frac{1}{m'}\left[2m|t'_0|\left(T^{-1}-\frac 13\epsilon T^{-\frac
13}\right)-\epsilon'T^{\frac 23}\right].
\eeq
This function is plotted in fig. \ref{Pperturb}. 

Note that this function diverges for  $T\lra 0$, obtains a minimum
for some small $T$ and increases thereafter. As $t'_0$ is driven
towards zero this minimum goes towards $T=0$. 

\section{The Gravitational entropy revisited}

Let us consider a finite 3-volume $V$ comoving with the cosmic gas in our space time. According to the 
first law of thermodynamics the matter entropy $S_M$ and the internal energy $U$ will evolve as:
\beq
TdS_M=dU+pdV.
\eeq
If the matter content is dust then 
$p=0$. 
There are a couple of things to note. Firstly, in a dense dust cloud, we expect
 the internal energy of the dust to be large, thus we expect the entropy to be 
large. Secondly, the entropy is an increasing function of the volume. 
Let us therefore consider a co-moving volume $V$ in our spacetime. What could 
the expression for a {\it gravitational entropy} be? From the previous 
sections we noticed that an LT model with $\bar{\rho}=\rho$ is homogeneous. 
The quantity $\frac{\bar{\rho}-\rho}{\rho}$ is an inhomogeneity measure in the
 LT models. In the absence of a cosmological constant we notice that 
\beq
P=\frac{2}{\sqrt{3}}\left(\frac{\bar{\rho}-\rho}{\rho}\right).
\label{eq:Pdiff}\eeq
The sign of $P$ is here chosen so that the  configuration 
$\bar{\rho}<\rho$ is associated with $P<0$ while the more realistic 
configuration $\bar{\rho}>\rho$ has $P>0$ corresponding to eq. (\ref{eq:Pdiff})
positive. Let us consider the entity defined by:
\beq 
{\mathcal S}=\int_V P dV.
\eeq
Introducing co-moving coordinates $x^i$ we write $dV=\sqrt{h}d^3x$ where 
$\sqrt{h}$ is the 3-volume element. The integration range is now constant as a
 function of time and if we integrate over a unit coordinate volume which is 
so small that the integrand is approximately constant, we may write 
\beq {\mathcal S}=\int_V P dV\approx P\sqrt{h}
\label{defS}\eeq

This formulation of the gravitational entropy is an intuitive one, but
it is
not a covariant formulation. To find a covariant formulation we define
the entropy current vector by \cite{hayward}
\beq
{\mbold \Psi}=s{\bf u}+{\mbold\varphi}
\eeq
where ${\bf u}$ and ${\mbold\varphi}$ are orthogonal, ${\bf s}$ is the
entropy density, ${\bf u}$ is the material flow vector (${\bf
u}\cdot{\bf u}=-1$) and ${\mbold\varphi}$ is the entropy flux. The second
law of thermodynamics can now be expressed as 
\beq
{\Psi^{\mu}}_{;\mu}\geq 0.
\eeq
Writing this in a local coordinate system, the divergence is
\beq
{\Psi^{\mu}}_{;\mu}=\frac{1}{\sqrt{|g|}}\partial_{\mu}\left(\sqrt{|g|}\Psi^{\mu}\right).
\eeq
In our case we take the entropy density proportional to $P$ and assume a
vanishing entropy flux. Hence,
\beq
{\mbold \Psi}\propto P{\bf u}.
\eeq
In comoving coordinates ${\bf u}={\bf e}_t$  so that 
\beq
{\Psi^{\mu}}_{;\mu}\propto\frac{1}{\sqrt{h}}\partial_t(\sqrt{h}P)=\frac{1}{\sqrt{h}}\dot{\mc{S}}.
\eeq
For the ``second law to
hold'', we have to check whether the entity $\mc{S}=\sqrt hP$ is
increasing. If we take the
corresponding one-form of the entropy vector, and  take the dual of
this form (Hodge dual), we  obtain the entropy  three-form which
is given by the contraction of the space-time volume form
${\mbold\epsilon}$ with the entropy current current vector
\beq
{\mbold\omega}=i_{\mbold\Psi}{\mbold\epsilon}.
\eeq
In our case, this three-form is simply
\beq
{\mbold\omega}=\sqrt{h}P{\bf dx}\wedge{\bf dy}\wedge{\bf dz}
\eeq
Hence, if the coordinates are comoving, then it is the component of
this three-form that is increasing as long at the matter and fields
obeys the SEC. This explains why the
entropy should scale as the volume. It is usually this entropy
three-form we are thinking of. It has a more intuitive behaviour than the entropy current vector.

We will therefore use $\mc{S}$ in the further to study the WCC. The square of the Weyl
tensor and the entity $P$ do not, as we have seen, capture the
entropic
behaviour properly. They both diverge at the initial singularity and
thus, cannot be a proper measure of the gravitational entropy. The
entity $\mc{S}$ on the other hand, seems to be a more promising candidate
for the Weyl entropy. 

\section{The behaviour of $\mc{S}$}
\subsection{The Lema\^itre-Tolman model}
Let us start with the LT model. The motivation for studying the
entity $\mc{S}$ given in (\ref{defS}) came from the LT models. Therefore it is natural to start with this model. In all
the LT models, we can approximate the behaviour near the initial
singularity, $R\lra 0$ with
\beq
\mc{S}_{LT}={2\sqrt{3}}\frac{m|t'_0|}{Fm'}\left(m'(t-t_0)-2t'_0\right).
\label{eqsmallt}\eeq
 For physically realistic spacetimes, $t'_0 < 0$, so as
$t\lra t'_0$, $\mc{S}$ is positive and finite. Note also that $\mc{S}$ is
increasing, as suggested by the WCC. In the absence of a cosmological
constant and if $F=1$, this will be the  exact expression for $\mc{S}$. If a
cosmological constant is present, the 
universe will eventually go into a de Sitter phase if the universe is
allowed to expand for ever.

\begin{figure}[tbp]
\centering
\epsfig{figure=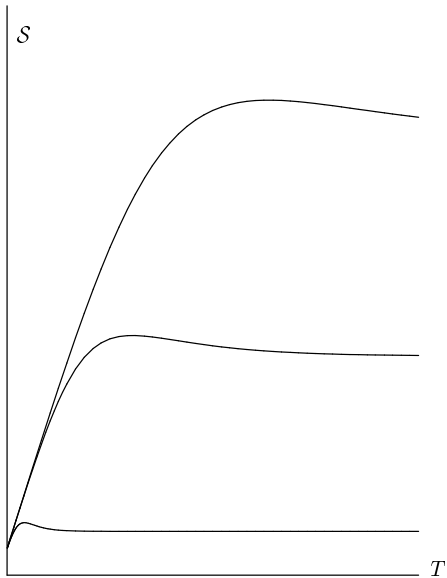, height=8cm}
\caption{The evolution of $\mc{S}$ in the LT models is dependent on the value of
$\Lambda$. The upper graph has a small value of $\Lambda$ which makes
$\mc{S}$ increase to a large value. The lowest graph has  a large value of
$\Lambda$.}
\label{fig:Sevol}
\end{figure}

For the sake of illustration, it is useful to investigate a specific
 case. Let us choose $F^2=1$. The energy equation
 eq. (\ref{energyeq}), can now be solved to yield
\beq
R=\left(\frac{6m}{\Lambda}\right)^{\frac 13}\sinh^{\frac
 23}\left[\frac 32 H(t-t_0(r))\right]
\label{eqExactLT}\eeq
where $H=\sqrt{\Lambda/3}$. In general, the late time behaviour of the
 LT model (if the universe has one) is that of a de Sitter
 universe. This exact solution provides us with a solution
 which connects the initial singularity with the late time de Sitter
 era. The initial behaviour of ${R}$ is the same as in eq. (\ref{smallimit}). The late
 time behaviour of $R$ is 
\beq
R=\left(\frac{3m}{2\Lambda}\right)^{\frac 13}e^{H(t-t_0(r))}.
\eeq
Hence, at late times the entity $\mc{S}$ is approximately constant:
\beq
\mc{S}=\frac{2m|t_0'|}{\Lambda^{\frac 12}}.
\label{eq:latetimedeSitter}\eeq
At late times, $\mc{S}$ approaches a constant value which is inversely
 proportional to $H$. The larger the value of the cosmological
 constant is, the
 smaller the value of the final value of $\mc{S}$. In
 fig. \ref{fig:Sevol} we have plotted, using the exact solution
 (\ref{eqExactLT}), the evolution of $\mc{S}$ for three different
 values of $\Lambda$. 

 The volume under consideration expands at late times 
exponentially as $V\propto V_0(r)\exp\left( 3Ht\right)$, where $H$ is
the Hubble parameter. Hence, the entropy per unit volume will decrease
exponentially:
\beq
\frac{\mc{S}}{V}\propto e^{-3Ht}.
\eeq 
The entropy in a unit volume will therefore  decrease rapidly
during a de Sitter stage. This is a very important consequence of
inflation. At the exit of the inflationary period, the entropy in a
unit volume is very low, there are very little inhomogeneities left of
the primordial ones. 

Let us summarize the evolution of the entity $\mc{S}$ in the LT models:
\begin{enumerate}
\item[$\bullet$]{\bf Large $\Lambda$ and ever expanding:} In the initial epoch
 the dust dominates and  $\mc{S}$ is increasing linearly in $t$. The universe is 
becoming more and more inhomogeneous. After the universe has grown  
considerably, the cosmological constant becomes dominant, $\mc{S}$ stops growing 
and if the cosmological constant is large enough, it evolves asymptotically 
towards a constant value. The universe is smoothened out. This
scenario corresponds approximately to the lowest two graphs in Fig. \ref{fig:Sevol}.
\item[$\bullet$]{\bf Small $\Lambda$ and ever expanding:} Again the dust 
dominates initially. The cosmological constant is too small to make $\mc{S}$ 
decreasing. The entity  $\mc{S}$ is ever increasing but is bounded from above by 
a relatively large constant value.
\item[$\bullet$]{\bf Zero $\Lambda$ and ever expanding:} The  $\mc{S}$ will again
 be ever increasing and will asymptotically approach a function 
$f(t)=c+bt^p$ where $c$ and $b$ are constants and $p=3$ iff $F^2>1$ and $p=1$ 
iff $F^2=1$.
\item[$\bullet$]{\bf Recollapsing universe:} Due to the dust term, the final 
singularity will not be similar to the initial singularity.   Hence this 
entity is asymmetric in time for a recollapsing universe.
\end{enumerate}

In the LT models we see that $\mc{S}$ behaves in agreement with the WCC. 

Let us investigate more carefully the Schwarzschild spacetime, where
the whole motivation of gravitational entropy comes from \cite{bekenstein,hawking}. The Schwarzschild spacetime is a
special case of the general LT models; it has $m(r)=constant$ and a vanishing Ricci 
tensor. If we look at the entity $\mc{S}$ in the region outside the Schwarzschild 
singularity, $\mc{S}$ will diverge. This is in some sense the maximal possible 
value of $\mc{S}$, the Weyl tensor is as large as possible and the Ricci tensor 
is the smallest as possible. Thus at this classical level it seems that the 
Schwarzschild spacetime has the largest possible $\mc{S}$ which is a good thing if 
one wants to connect $\mc{S}$ with the entropy of the gravitational
field. Outside a real black hole, $R_{\mu\nu}=0$ is probably
impossible. Even though the classical vacuum has vanishing Ricci
tensor, the quantum vacuum will probably not have $R_{\mu\nu}=0$. The
quantum fields will fluctuate and cause the expectation value of the square of the Ricci
tensor to be non-zero: $\left\langle R_{\mu\nu}R^{\mu\nu}\right\rangle
\neq 0$. Hence, there will probably be an upper bound of how large
$\mc{S}$ can be, even in a vacuum. 

\subsection{The Bianchi type I model}
In the Bianchi type I with dust, the explicit expression for the entity
$\mc{S}$ is 
\beq
\mc{S}_I=\sqrt{h}P=\frac{2A^2}{3\sqrt{3}}\left(\frac{1+z^2-2z\cos3\gamma}{M^2+
2\Lambda Mv+4\Lambda^2v^2}\right)^{\frac{1}{2}}
\eeq
where $\sqrt{h}=v$. In a neighbourhood of the singularity we can make a
Taylor expansion to first order in $v$:
\[
\mc{S}^2_I\approx \frac{8}{27}\frac{A^4}{M^2}(1-\cos3\gamma)\left(1+
\left(\frac{3M}{2A^2}-\frac{2\Lambda}{M}\right)v\right).
\]
We interpret $m=M/A$ as the dust density in coordinate space
\cite{sigBI}. Hence, as we go towards the initial singularity $\mc{S}$
will be finite. For the vacuum case, $m=0$ and $\Lambda=0$, and the
solutions are the Kasner solutions. The Kasner solutions have a
vanishing Ricci tensor, but a non-zero Weyl tensor. Hence, as in
the Schwarzschild case, the entity $\mc{S}$ will diverge. 

From the Taylor expansion we see that, if $3m^2>4\Lambda$ then $\mc{S}$
will increase immediately after the initial singularity. If
$3m^2<4\Lambda$ then $\mc{S}$ will decrease. This can be understood as
follows. If the cosmological constant is to large, the universe will
increase too rapidly initially to allow the dust to contribute
significantly to the anisotropy. Initially the universe will become more and
more isotropic. In the presence of a cosmological constant, the
universe will eventually enter a de Sitter phase. Again $\mc{S}$ will evolve
towards a constant value given by
\[
\lim_{v\ra\infty}\mc{S}_I=\frac{A}{3\Lambda^{\frac{1}{2}}}.
\]

If the cosmological constant vanishes $\mc{S}$ is
monotonically  increasing:
\[
\mc{S}_I=\frac{2A}{3\sqrt{3}m}\left((2+3mt)(1-\cos 3\gamma)+
\frac{9}{4}m^2t^2\right)^{\frac{1}{2}}
\]
Note that for large $t$ we have $\mc{S}_I=\frac{A}{\sqrt{3}}t$, which is
according to the same power law as the $F^2=1$ and $\Lambda=0$ case of
the LT model (compare with eq. (\ref{eqsmallt})). 

To summarize our investigation of $\mc{S}$ in the Bianchi type I model we can say
 that the entity $\mc{S}$ at the initial singularity is a constant determined by 
the inverse of the dust density. For $3m^2>4\Lambda$ it will increase 
immediately after the initial singularity. In most cosmological considerations
 it is assumed that the cosmological constant is small. The exception is in 
the inflationary era in which the vacuum energy dominates over  all other 
matter degrees of freedom. The inflationary era will smooth out anisotropies 
as well as inhomogeneities, and the behavior of $\mc{S}$ in this case is 
therefore expected. In the de Sitter limit $\mc{S}$ will asymptotically move 
towards a constant. Large $\Lambda$ means small value, while small $\Lambda$ 
corresponds to a large value. This is in full agreement with the LT models. 
It is also interesting that the entity $\mc{S}$ is very sensitive to different 
matter configurations. This makes it a lot easier to check whether the $\mc{S}$ has the right entropic behaviour. 

The question now arises: How generic is this behaviour? Does the
entity behave in the correct way for all physically realistic models?

We know that our universe today is close to homogeneous on a scale
larger than a billion light years. As
mentioned earlier, the FRW models are conformally flat and hence, they
have a zero Weyl tensor. So a question would be, at late times when
the universe is close to isotropic and homogeneous, does the entity
$\mc{S}$ still behave in the correct manner? Does it still increase? We
saw  that at late times, both the LT model and the Bianchi type I
model, did behave correctly for $\Lambda=0$ even though they both
isotropise and evolve towards homogeneity. 

Let us choose a more general model, a flat model that allows for
inhomogeneities and anisotropies. We will assume that the model at
late times asymptotically evolve towards a FRW model. In
\cite{BM,Barrow}, Barrow and Maartens investigated the general
equations of motion for such models with $\Lambda=0$. In the velocity dependent regime
they derived some approximate solutions to the field equation close to a FRW
model, having both inhomogeneities and anisotropy. Also an
anisotropic stress of the form $\pi_{\mu\nu}=\lambda_{\mu\nu}\rho_r$
where $\lambda_{\mu\nu}$ is a constant matrix, was included. Their result was that the late time
behaviour of such a model is consistent with a perfect fluid where the
equation of state parameter $\gamma$ for the perfect fluid obeys
$\gamma \leq 4/3$. For $\gamma > 4/3$ the FRW model is unstable at late
times. In the latter case one can get for instance different stable anisotropic
magnetic solutions
\cite{Jacobs,LeBlanc}. 

Let us therefore assume that $2/3 \leq \gamma\leq 4/3$ (so that the
matter obeys the SEC). 
The ratio of the Weyl tensor squared and the Ricci tensor squared is
in this case
\beq
P^2=\frac{C^{\alpha\beta\gamma\delta}C_{\alpha\beta\gamma\delta}}{R^{\mu\nu}R_{\mu\nu}}\propto
\mc{O}(t^{-2n})+\mc{O}(t^{-3n})+\mc{O}(t^{-4n})
\eeq
where $n=\frac{2(4-3\gamma)}{3\gamma}$ for $n \neq 4/3$. In the
radiation case, $\gamma =4/3$, we get a logarithmic decay of $P^2$:
\beq
P^2_{\gamma=4/3}\propto \frac{1}{(\ln t)^2}.
\eeq 
Thus in the range of validity of the assumption $\gamma \leq\frac 43$,
this entity will decrease in the future. However, the entity $\mc{S}$ shows more promising behaviour
\beq
\mc{S}=P\sqrt{h}=a^3 P\propto t^{\frac{2}{\gamma}-n}
\label{eq:BMsol}\eeq
for $2/3 \leq \gamma < 4/3$, while the $\gamma =4/3$ case yields
\beq
\mc{S}\propto (\ln t)^{-1}t^{\frac 32}.
\eeq
Hence, $\mc{S}$ increases as long as
\beq
\frac 23\leq \gamma\leq \frac{4}{3}.
\eeq
This entity increases in the future (as any entity describing
entropy should do) and tells us that even if the universe itself
asymptotically goes towards isotropy, the entropy of the gravitational
field actually increases (if we should believe the WCC). 

Note that this result is slightly different than the case where no
anisotropic stress is present. If the anisotropic stress is not
present, we get $\mc{S}\propto t$ at late times for all $\gamma$. 

We should also mention a work done by Hervik \cite{hervik} which investigates the
evolution of the 
Weyl curvature invariant for generic solutions of spatially homogeneous models
containing a $\gamma$-law perfect fluid with $\gamma\geq 2/3$. The conclusion was that all
spatially homogeneous  models, except for sets of measure zero, had an increasing
$\mc{S}$ at late times. 

\subsection{Inflation}

As long as the matter obeys the strong energy condition (SEC), matter will
behave more or less attractive. Hence, when the SEC is fulfilled, 
we should expect the gravitational entropy to increase.

 During inflation the SEC is
violated and gravity is not necessarily attractive. We have already seen
how the inclusion of a cosmological constant could alter the behaviour
of $\mc{S}$. The late time behaviour of $\mc{S}$ in the presence of a
cosmological constant is that $\mc{S}$ evolves approximately as a
constant. The constant itself is a decreasing function of
$\Lambda$ (see eq. (\ref{eq:latetimedeSitter})). 

What happens for a more general inflationary fluid? Consider a perfect
fluid with equation of state 
\beq
p=(\gamma-1)\rho
\eeq
for $0\leq\gamma<\frac{2}{3}$. This perfect fluid will violate the strong
energy condition and will in general cause a power law inflation. Note
that the 
case $\gamma=0$ can be considered the same as including a cosmological
constant. In the FRW cases, all the models will now have a late time
behaviour similar to the flat case. The FRW models will become
inflationary and will be dominated by this fluid at some stage in the
future (provided that the universe is ever-expanding). To simplify,
we will therefore consider a flat universe, and perturb the FRW flat
universe model with
$\gamma <2/3$. Following \cite{BM,Barrow}, the anisotropic stresses
will not dominate the shear modes at late times. The shear will under these assumptions decrease as
\beq
\sigma_{\mu\nu}\propto t^{-\frac{2}{\gamma}}
\eeq
while the Hubble parameter is the same as in the FRW case (to lowest
order)
\beq
H=\frac{2}{3\gamma}t^{-1}.
\eeq
The late time behaviour of $P^2$ is now 
\beq
P^2= \mc{O}\left(\gamma^2 t^{-\frac{2(2-\gamma)}{\gamma}}\right)
\eeq
which is decreasing for all $0<\gamma<2/3$. The late time behaviour of
$\mc{S}$ is 
\beq
\mc{S}=\sqrt{h}P=\mc{O}\left(\gamma t\right)
\eeq
which is increasing linearly in $t$ at late times. For $\gamma=0$ we
recover the cosmological constant case where $\mc{S}$ is approximately
a  constant at late times. Note also that this result coincide with
eq.  (\ref{eq:BMsol}) for $\gamma =2/3$. 

Hence, during the inflationary period, $\mc{S}$ increases much slower
than for ordinary matter ( $\mc{S}\propto t^{\frac 43}$ for dust and
$\mc{S}\propto (\ln t)^{-1}t^{\frac 32}$ for radiation).

\section{Using Quantum Cosmology to determine the initial state}
Maybe the closest to a quantum theory of gravitation that have been obtained, is
what we call quantum cosmology (QC). We will in this section show how one might be able to determine the likelihood of a certain initial
state to occur. QC is perhaps best described as a ``theory of initial
conditions''. We will again use the results of the previous paper to
try 
determine which of the  initial states are more probable.

In QC, the wave function of the universe satisfies the
\emph{Wheeler-DeWitt (WD) equation}:
\beq
\left(-G_{ijkl}\frac{\delta}{\delta h_{ij}}\frac{\delta}{\delta
h_{kl}}+ V(h_{ij})\right)\Psi =0
\eeq
where $G_{ijkl}$ is called DeWitt's supermetric, and $h_{ij}$ is the
metric on the 3-dimensional spatial hypersurfaces. The potential term
$V(h_{ij})$ consists of the Ricci scalar of the three-dimensional
hypersurfaces and possibly matter potentials and a cosmological
constant. 

We want now to calculate the expectation value for the Weyl scalar, in
these models. All of the curvature invariants have to go over to their
respective curvature operators. Especially, the entity $\mc{S}$ goes over
to the curvature operator $\widehat{\mc{S}}$. The question is now, what
is more likely, creation of a universe with large expectation value
of $\widehat{\mc{S}}$ or a universe with a low value? 

\subsection{The LT model}

In the paper \cite{sigLT} we considered 
semi-classical tunneling wave functions which are solutions of the WD
equation. The universe was tunneling from a 
matter-dominated universe classically confined to a finite size, 
into a de Sitter like universe. After tunneling across the classically 
forbidden region the universe became $\Lambda$-dominated, similarly to an
 inflationary model of the universe. Thus we have a ``flow'' from
matter-dominated universes towards de Sitter like universes.  

The actual expectation values of entities like 
$C^{\alpha\beta\gamma\delta}C_{\alpha\beta\gamma\delta}$, $P^2$ and $\mc{S}$ are 
not explicitly obtained for these models, because the actual calculations 
suffer from ``endless'' expressions and highly time-consuming quantities. We 
will therefore give a more general description of the evolution of the Weyl 
tensor for the LT models. 

It would be useful first to recapitulate some of the discussion done in 
\cite{sigLT,wcc1}. First of all we discussed tunneling wave functions in the WKB 
approximation. In the WKB approximation we assume that the wave function has 
the form 
$\Psi_{WKB}=\exp(\pm iS)$,
where $\mc{S}$ will to the lowest order satisfy the Hamilton-Jacobi equation: 
\begin{equation}\label{energidiff}
\left(\frac{\delta S}{\delta R}\right)^2-
\frac{{F'}^2}{F^4}\left[2mR-R^2(1-F^2)+\frac{\Lambda}{3}R^4\right]=0
\end{equation}
Here $\delta/\delta R$ denotes the functional derivative with respect
to the function $R$. 
If we assume that $S=\int \sigma(r) dr$, the resulting equation will be  
the Hamilton-Jacobi equation for a point particle with action $\sigma(r)$
($r$ is only a parameter and the functional derivatives turn into
ordinary partial derivatives). In the Hamilton-Jacobi equation the functional 
$S$ turns out to be the action at the classical level. Since the classical
 action can be written as an integral over $r$ the assumption 
$S=\int \sigma(r) dr$ is therefore reasonable at the lowest order WKB 
level. 
We can interpret the action $\sigma$  as the action of a point particle 
moving in a potential 
$V(R)=\frac{{F'}^2}{F^4}\left[-mR
+\frac{1}{2}(1-F^2)R^2-\frac{\Lambda}{6}R^4\right]$ with zero energy. The 
WKB wave function $\psi_{WKB}$ for the point particle can then be written
$\psi_{WKB}=\exp(\pm i\sigma)$. The two WKB wave functions can therefore 
be 
related by $\Psi_{WKB}=\exp(\int dr\ln \psi_{WKB})$. Finding first the 
wave function $\psi$ we can then relate its WKB approximation to
$\Psi_{WKB}$ through  $\Psi_{WKB}=\exp(\int dr\ln \psi_{WKB})$.

Let us now ask the question: \emph{Is it more likely for a universe with small 
Weyl tensor to tunnel through the classical barrier than a universe with a 
large Weyl tensor?} The question is difficult to answer in general but we shall make
 some simple considerations in order to shed some light upon it. 

We assume that the dust density near the origin of the coordinates is larger 
than further out. We define the homogeneous mass function for a closed
universe ($k=1$) as 
$m_h(r)=\frac{4}{3}\pi\rho_h r^3$ where $\rho_h$ is a constant. The
constant $\rho_h$ is determined by demanding
$m_h(r_{max})=m(r_{max})$. If the 
dust density is larger near the origin of the $r$-coordinate than for larger 
values of $r$ then $m(r)\geq m_h(r)$. This will not in general 
change the size of the universe so we can look at the effects from $m(r)$ 
alone. 
Since $m(r)$ is greater in general for an inhomogeneous universe than for a 
homogeneous universe, we see that the potential barrier will be smaller for an
 inhomogeneous universe than for a homogeneous universe. Thus an 
inhomogeneous universe will tunnel more easily through the classical barrier 
than the homogeneous universe. Since an inhomogeneous universe will have a 
larger Weyl tensor than an almost homogeneous one, we see that universes with 
large Weyl tensor tunnel more easily than those with a small Weyl tensor. 
  
If we look at the tunneling amplitude concerning effects from the $\Lambda$ 
term, it is evident that larger $\Lambda$ will yield a larger tunneling 
probability.  In the initial era inhomogeneities will increase the value of 
$\mc{S}$. We saw that  an inhomogeneous state will tunnel more easily through the
 potential barrier than a homogeneous state.  The largest tunneling 
probability amplitude thus occurs for universes with a large cosmological 
constant and large local inhomogeneities. From a classical point of view the 
value of $\mc{S}$ initially was large (but increasing thereafter), but as the 
universe entered the inflationary era the cosmological constant had a
large value, hence the value of $\mc{S}$ at the end of the
inflationary era was  relatively  
small. 

In the initial epoch the universe is not believed to be dust dominated. The 
dust does not exert any pressure and dust particles do therefore not interact 
with each other. A more probable matter content is matter which has internal 
pressure. Even though gravitation tends to make the space inhomogeneous, 
internal pressure from the matter will try to homogenise the space.  Since 
dust is the only  matter source in our model, the model only indicates the 
tendency for gravitation itself to create inhomogeneities. 

Since the universes tunnel into a de Sitter-like state, the cosmological 
constant will rapidly dominate the evolution. The larger the cosmological 
constant the lower will the entity $\mc{S}$ be after the inflationary era ends.

\subsection{The Bianchi type I model}

We can write the general solution of the WD-equation for the Bianchi type I models as:
\begin{equation}
\Psi(v,\beta)=\int d^2k\left[C(\vec{k})\psi_{\vec{k}}(v)\rho(\vec{k})
e^{i\vec{\beta}\cdot\vec{k}}\right]
\end{equation}
where $\psi_{\vec{k}}(v)=v^{-\frac{\zeta}{2}}W_{L,\mu}(2Hv)$ is a particular 
solution of the WD equation (with dust), $C(\vec{k})$ is a ``normalizing 
constant'', and $\rho(\vec{k})$ is a distribution function in momentum space. 
This distribution function satisfies the equation:
\[ \int d^2k |\rho(\vec{k})|^2=1 \]  
The function $W_{L,\mu}(x)$ is the \emph{Whittaker function}.

There is still a factoring problem in turning the classical entities to 
operators. Let us first investigate the expectation value of 
$C^{\alpha\beta\gamma\delta}C_{\alpha\beta\gamma\delta}$. In the
expression for the Weyl tensor squared, there is a term 
\beq
\frac{1}{v^4}\left(p_+^2+p_-^2\right)^2
\eeq
where $p_{\pm}$ is the conjugated momenta to $\beta_{\pm}$. Upon
quantization, the above term is replaced with the operator:
\beq
\frac{1}{v^4}\left(p_+^2+p_-^2\right)^2\longmapsto \frac{1}{v^4}\left(\frac{\partial^2}{\partial \beta_{+}^2}+
\frac{\partial^2}{\partial \beta_{-}^2}\right)^2
\eeq
In the expectation value of the square of the Weyl tensor, the above
term will contribute with
\[ \frac{\int d^2\beta \Psi^*\frac{1}{v^4}\left(\frac{\partial^2}{\partial 
\beta_{+}^2}+
\frac{\partial^2}{\partial \beta_{-}^2}\right)^2\Psi}{\int d^2\beta \Psi^* \Psi}
 = \frac{1}{v^4}\frac{\int d^2k|C(\vec{k})|^2|\psi_{\vec{k}}(v)|^2
|\rho(\vec{k})|^2|\vec{k}|^4}{\int d^2 k|C(\vec{k})|^2|\psi_{\vec{k}}(v)|^2
|\rho(\vec{k})|^2}.
\]
Unless we have a delta-function distribution at $\vec{k}=0$; 
$\rho(\vec{k})=\delta^2(\vec{k})$, the contribution from this term to the 
Weyl curvature invariant will diverge as $v^{-4}$ for small $v$. Thus, we have to conclude 
that, in the small $v$ limit the expectation value of 
$C^{\alpha \beta \gamma \delta}C_{\alpha \beta \gamma \delta}$ goes as:
\begin{equation}
\inl C^{\alpha \beta \gamma \delta}C_{\alpha \beta \gamma \delta}\inr\propto 
\frac{1}{v^4}
\end{equation} 
just as in the classical case. 
\par
Investigating the invariant $R^{\mu \nu}{R_{\mu \nu}}$, we notice that things 
are not so easy. The Ricci square also has a term which presumably would 
contribute with a $\frac{k^4}{v^4}$ term. However, looking at the classical 
expression we see that the Ricci square is independent of the anisotropy 
parameter $A$. This indicates that at the classical level all terms involving 
the anisotropy parameters, have to cancel exactly. This is not the case 
quantum mechanically. In the quantum case operators do not necessarily 
commute. Hence there may be contributions from  terms which classically would 
cancel each other. In other words, the fact that the classical vacuum has 
$R_{\mu\nu}=0$, does not mean that the quantum vacuum has $\hat{R}_{\mu\nu}=0$.

We assume that $(\xi_j)$ is a set of factor-ordering parameters which 
represents the ``true'' quantum mechanical system in such a way that  
$\xi_j=0$ represents the classical system. With this parameterization of the 
factor ordering we would expect the Ricci square expectation value to be:
\[ \inl R^{\mu \nu}{R_{\mu \nu}}\inr=4\Lambda^2+2\Lambda\frac{M}{v}+
\frac{M^2}{v^2}+\frac{f_j(v)}{v^4}\xi_j+{\mathcal{O}}(\xi_i^2) \]
where $f_i$ is some function of $v$ which has the property: 
$v\approx 0, \quad f_j(v)\approx constant$.
Thus for small $v$ and $\xi_j$ the expectation value would behave as
\[  \inl R^{\mu \nu}{R_{\mu \nu}}\inr \propto \frac{f_j(0)}{v^4}\xi_j \]

and
\begin{equation}
\frac{\inl C^{\alpha \beta \gamma \delta}C_{\alpha \beta \gamma \delta}\inr}
{\inl R^{\mu \nu}{R_{\mu \nu}}\inr} \propto\frac{1}{f_j(0)\xi_j}\cdot constant
\end{equation}
The Weyl square divided by the Ricci square is in general finite as 
$v\longrightarrow 0$ for a quantum system. In some sense, the quantum
mechanical effects renormalizes the infinity that the classical
system possesses at $v=0$. The expectation value at $v=0$ is,
however, strongly dependent on the factor-ordering. As the factor ordering 
parameters approach zero, the value will diverge. Quantum effects in the early
 epoch are essential for the behaviour of this entity near the initial 
singularity. As $v\longrightarrow 0$ we expect the quantum effect to be 
considerable, thus expecting the factor-ordering parameters $\xi_j$ to be
 large.

As indicated in the above discussion, the quantum mechanical expectation value of
 $\hat{P}$ will be lower and presumably finite at the initial singularity. 
Therefore the expectation value of $\hat{\mc{S}}$ is also presumed to be 
considerably lower in the initial stages than its classical counterpart. 

Comparing different tunnelling amplitudes in the Bianchi type I model is 
difficult and more speculative because the Bianchi type I universe has no 
classically forbidden region for $\Lambda\geq 0$. This causes the lowest order
 WKB approximation to be purely oscillatory. The lowest order WKB wave function
 will therefore be approximately constant. In the paper \cite{sigBI} we did 
however construct under some assumptions a wave function which clearly peaked 
at small values of the anisotropy parameter. Thus these wave functions predicts
 universes that have a relatively low value of $\mc{S}$. 

\section{Cosmic evolution of the Weyl Curvature}
Let us now recapitulate how the evolution in the context of the WCC
\emph{might} have been. 

The universe was created in a rather arbitrary state. As the time
ticked past the Planck time $10^{-43}$ s a rather inhomogeneous
universe appeared. QC
suggests that this state was rather inhomogeneous, but it is more or
less a guess how inhomogeneous the universe was at that
state. Nevertheless, as the universe grew larger, the expectation value
of the Weyl entropy
increased initially. At what rate the Weyl entropy increased is highly
uncertain, it depends very much on the true nature of our universe. It depends on
what matter fields that were present, the topology of the universe,
whether it was anisotropic or not, quantum effects etc. 

Nevertheless, at some time very short after the Big Bang, an enormous effective
cosmological constant appeared. The universe was driven unconditionally into an inflationary
period. The Weyl entropy stopped increasing and began instead to
evolve asymptotically towards a constant. Whether or not the Weyl entropy had earlier
a higher value than at the exit of the inflationary regime, is
difficult to say. During the inflationary period, the 
universe was more or less in an adiabatic expanding state, the Weyl entropy
was more or less constant. Since the universe was expanding
exponentially during this period, the entropy per unit volume dropped
exponentially. If the scale factor increased by a factor of 60
$e$-foldings during the inflationary epoch, then the entropy per unit
volume of space  would have decreased by a factor of
\beq
\frac{s}{s_0}\sim e^{-3\cdot 60}\approx 10^{-78}.
\eeq
This is quite a drastic decrease, and shows how powerful inflation is when it
comes to smoothing out the inhomogeneities and anisotropies of our
universe. 

\begin{figure}[tbp]
\centering
\epsfig{figure=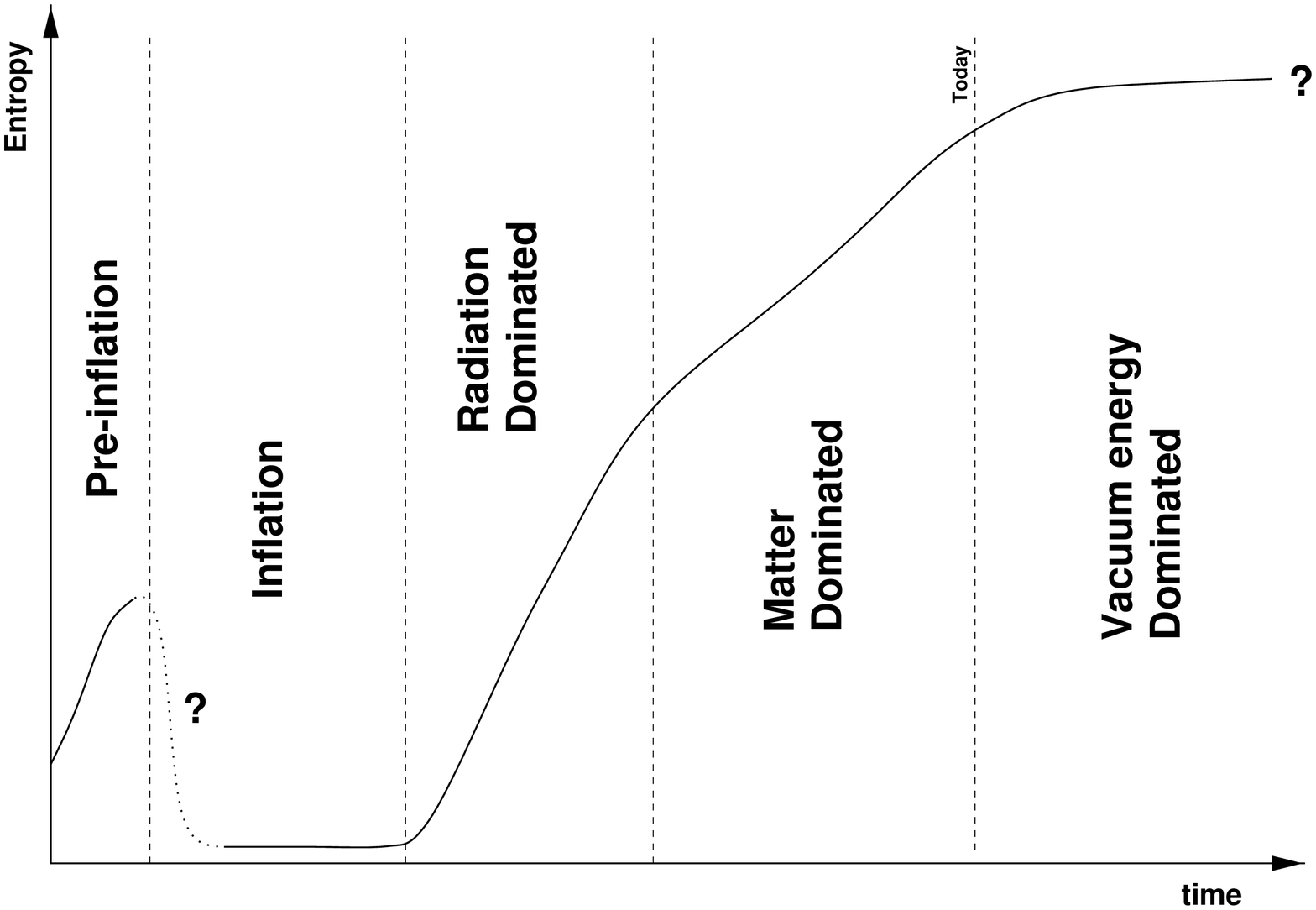, width=10cm}
\caption{A schematic graph of the evolution of the Weyl entropy. }
\label{fig:weylevol}
\end{figure}

At the exit of the inflationary regime, the Weyl entropy had dropped
by an enormous factor compared to what is would have been if no
inflation had occurred. The Weyl entropy was very small, compared to
the maximal possible value. The universe was more or less uniform and
homogeneous. This homogeneity can be seen in the CMB radiation
today. However, small fluctuations in the spectrum can also be seen,
reflecting the state of the universe 300 000 years after the Big
Bang. After the inflation, small seeds of inhomogeneities from the
quantum fluctuations of quantum fields were the only thing left of
the initial inhomogeneities. Nevertheless, these seeds were large enough to
gradually clump together and form galaxies and stars. The Weyl entropy
began to increase again after the inflationary regime was over. 

The radiation-, and later the matter-dominated universe caused the
Weyl entropy to grow steadily and firmly for almost 15 billion
years. During the radiation era, there were only small inhomogeneities
and anisotropies left of the primordial fluctuations. Using the
calculations from the earlier section, the Weyl entropy increases as
$\mc{S}\propto (\ln t)^{-1}t^{\frac 32}$ in the radiation era. 

At about $t=$10 000 years  the
radiation become sub-dominant. The universe evolved into a matter
dominated era and the Weyl entropy increased as $\mc{S}\propto t^{\frac
43}$. Today it is still  growing. Recent observations 
suggest that the universe has entered a new era with accelerated expansion. The surprising fact
that the universe appears to be in an accelerating state today, can be
explained with the presence of a  vacuum energy. If
this is true, the Weyl entropy will increase steadily and
asymptotically towards a constant. However, the late time behaviour of
our universe is still quite speculative, and whether or not this
vacuum dominated period will persist, is very uncertain. It might
happen that the period ends like inflation did, and perhaps the vacuum
period is followed by a curvature dominated period. If so, the Weyl
entropy might again rise to new heights, increasing towards a value
where all the matter is collected in black holes.

\section*{Acknowledgments}
SH was founded by the Research Council of Norway.

\end{document}